\documentclass[12pt,leqno,a4paper]{article} 
\usepackage{epsfig}
\usepackage[english]{babel}

\begin{document}

\title{Jets in Deep Inelastic Scattering and High Energy Photoproduction at
HERA \footnote{Talk given at the 9th Adriatic Meeting ``Particle Physics and the
Universe'' in Dubrovnik/Croatia, 4.-24. September 2003}}

\author{Gerd W. Buschhorn \\ Max-Planck-Institut f\"ur
Physik \\(Werner-Heisenberg-Institut) \\F\"ohringer Ring 6 \\80805 M\"unchen} 
\date{}

\maketitle

{\abstract Recent results on jet production in neutral current deep inelastic
scattering and high energy photoproduction at the HERA electron-proton-collider
are briefly reviewed. The results are compared to QCD expectations in NLO and
$\alpha_s$ determinations using these data are summarized.}

\section{Introduction}
\label{sec:1}

The hard partonic interactions at the center of photon-induced hadronic
processes in deep inelastic electron/positron-proton-scattering are
charaterised by jets. The virtuality $Q^2$ of the exchanged photon can be
varied from high values in deep inelastic scattering (DIS) to zero in
photoproduction. According to the factorization theorem the hadronic process
can be factorized into a pertubatively treated hard part, which includes the
evolution of the parton entering the hard process,  and a soft part, which
describes the parton distribution of the target nucleon; the characteristic
energy for this separation is given by the factorization scale. In DIS
processes involving light quarks only, an energy scale is provided by $Q^2$ or
the transverse energy of the emerging hadronic jet (or jet system), whereas in
photoproduction the jet energy is the only available energy scale; in heavy
quark production the quark mass may provide an additional energy scale. In jet
studies details of the hard QCD calculations and their matching to the
evolution of the target partons are probed; the investigations may also provide
information on the parton distribution functions (pdfs) of the proton and
photon and can serve to measure $\alpha_s$ and its scale dependence.

In this report, recent results from the H1- and ZEUS-collaboration on the
production of standard jets i.e. jets without heavy quark tagging in neutral
current DIS processes and photoproduction are discussed; jets from diffractive
processes and jet shapes are not included.

For jet studies in DIS the Breit frame is preferred, which in the quark parton
model (QPM) is defined as the reference frame in which the struck parton
emerges collinear with the incoming purely time-like virtual photon
$\mathbf{q}=(0, 0, -Q)$ such that $2x\mathbf{p+q} = 0$. Working in the Breit
frame suppresses the QPM background and provides maximum separation between
fragments from hard scattering QCD processes and beam remnants. Since the Breit
frame is related to the photon-hadron-cms by a boost in proton beam direction
(defined as $+z$-direction), jet variables are convenient, which are
boost-invariant - like the transverse energy $E_T$, the azimuthal angle $\phi$
and the pseudorapidity $\eta = -\ln tan (\theta/2)$.

The jet finding algorithm has to be stable against infrared radiation and
collinear splitting and has to take into account the fact that at HERA not all
final state particles are detected. In recent years, the algorithm preferred in
jet studies by the HERA-collaborations has become the inclusive $k_{\rm
T}$-cluster algorithm \cite{ref1}. The algorithms proceeds by calculating for
each track or energy cluster $i$ the quantity $d_i = (E_{Ti})^2$ and for each
pair $i,j$ the quantity $d_{ij} = \min (E^2_{Ti}, E^2_{Tj})[(\eta_i - \eta_j)^2
+ (\phi_i - \phi_j)^2]$. If of all resulting $\{d_i, d_{ij}\}$ the smallest one
belongs to $\{d_i\}$ it is kept as a jet and not treated further, if it belongs
to $\{d_{ij}\}$ the corresponding tracks or clusters are combined to a single
object; the procedure is repeated until all tracks or clusters are assigned to
jets. The algorithm is suited to handle overlapping jets, facilitates the
separation between beam remnants and hard scattering fragments and yields
smaller higher order QCD and fragmentation corrections than e.g. the cone
algorithm.

\begin{figure}
\center
\includegraphics[height=5cm]{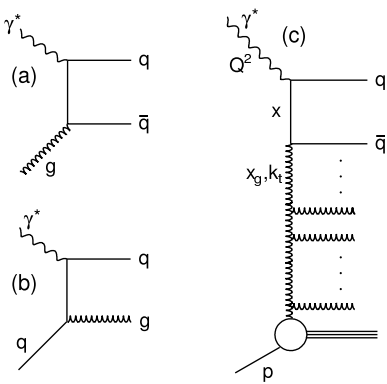}
\caption{\small Leading order diagrams for jet production in photon-proton-interaction:
(a) photon-gluon-fusion; (b) QCD-Compton process; (c) parton evolution}
\label{fig1}
\end{figure}

Jet cross sections are calculated by convoluting the matrix element of the
specific hard scattering process with the pdfs of the target proton. Examples
of LO hard scattering processes are boson-gluon-fusion and QCD-Compton
scattering (fig.~\ref{fig1}). The parton evolution requires the summation of
$\ln Q^2$- and $\ln 1/x$-terms, which is equivalent to a summing of ladder
graphs pictured in fig.~\ref{fig1}. In the DGLAP-approach \cite{ref2} only
leading powers of $\ln Q^2$-terms are summed implying a strongly ordered
increase of the virtuality of the exchanged partons towards the hard
scattering. At small $x$ where at HERA $Q^2$ is constrained to not too large
values, the BFKL treatment of the parton evolution \cite{ref3} is more
appropriate, which resums  $\ln 1/x$-terms to all orders but retains the full
$Q^2$ dependence; the parton chain is strongly ordered in $x$ i.e. $x \ll x_g
\ll ... \ll 1$, implying $k_{\rm T}$-unintegrated (gluon) pdfs and off-shell
matrix elements for the hard scattering process. In the CCFM ansatz \cite{ref4}
colour coherence in the evolution ladder is taken into account imposing an
angular-ordering of the gluon emission with the maximum angle determined by the
hard scattering subprocess; the cross section factorizes into pdfs unintegrated
both in $k_{\rm T}$ and the gluon angle and an off-shell matrix element. At
asymptotic energies CCFM evolution approaches BFKL at small $x$ and DGLAP at
large $x$ and $Q^2$.

While at high $Q^2$ the virtual photon behaves as a pointlike particle, at low
$Q^2$ and in particular in photoproduction with $Q^2 = 0$, its hadronic structure
has to be taken into account. The momentum transfer from the scattered electron
to the parton cascade then takes place via a parton of the "resolved photon",
which itself may evolve in a parton cascade towards the hard scattering process.
Such resolved processes contribute a background to hard scattering processes but
also can serve as a tool to investigate the photon structure function.

The response of the detectors to jet events is studied with Monte Carlo
generated events, which are based on the measured pdfs of the proton and photon
and on simulations of the fragmentation and hadronization of the partons. In
RAPGAP the QCD matrix element is calculated in LO with NLO corrections evaluated
by perturbative generation of parton showers; the hadronization of the partons
is performed via JETSET, which is based on the Lund-string model. In ARIADNE,
the parton cascade is generated using the colour-dipole-model (CDM) while in
CASCADE it is calculated via CCFM evolution.

\section{Jets in DIS}
\label{sec:2}

The inclusive jet cross section $\sigma_j$ results from the convolution of the
measured pdfs of the proton $f_a (x_a, \mu_F)$ with the pertubatively
calculated hard partonic cross section $d\sigma_a (x_a, \alpha_s, \mu_R, \mu_F)$,
\[ 
\sigma_j = \sum_a \int dxf_a(x,\mu^2_F)
d\sigma_a(x,\alpha_s,\mu^2_R,\mu^2_F)(1+\delta_{\rm {hadr}})
\]

\noindent where the sum is to be taken over all partons $a$ of the proton;
$\sigma_j$ has to be folded with the bremsstrahlungsspectrum. $\mu_R$ is the
renormalization scale and $\mu_F$ the factorization scale. $(1+\delta_{\rm
{hadr}})$ represents the nonperturbative correction accounting for the
hadronization of the partons; it is estimated from MC generators.

\begin{figure}
\center
\includegraphics[height=7cm]{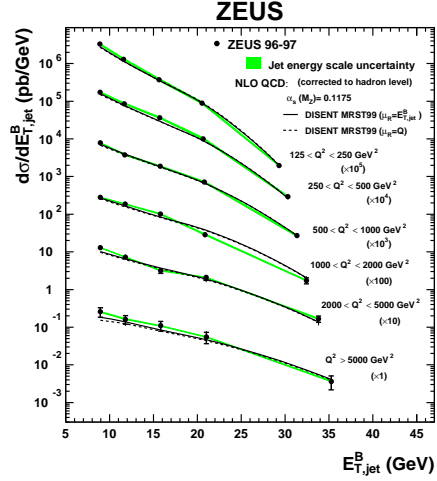}
\caption{\small Inclusive jet cross section in DIS in the Breit system (fig. from
\cite{ref6})}
\label{fig2}
\end{figure}

\subparagraph{(i)} Single jet cross sections have been measured by H1 and ZEUS
in the Breit frame using the $k_{\rm T}$-cluster algorithm. The kinematical
parameters for the H1 data \cite{ref5} are $150 < Q^2 < 5000$ GeV$^2$, $7 < E_T
< 50$ GeV, $-1 < \eta_{\rm {lab}} < 2.5$, while for the ZEUS data \cite{ref6}
they are $Q^2 > 125$ GeV$^2$, $E_T > 8$ GeV, $-2 < \eta_{\rm lab} < 1.8$. The
good agreement of both measurements with NLO QCD calculations (DISENT) at high
$Q^2$ values (fig.~\ref{fig2}) has enabled precision determinations of $\alpha_s$
and its scale dependence from these data (see Sect. \ref{sec:5}).

The ZEUS data have been used to study the azimuthal distribution of jets in the
Breit frame \cite{ref7}. The measured $\phi$-distribution agrees well with the
NLO QCD predictions from DISENT with either $\mu_R = E_T$ or $Q$, providing an
independent test of QCD.

\begin{figure}
\center
\includegraphics[height=7cm]{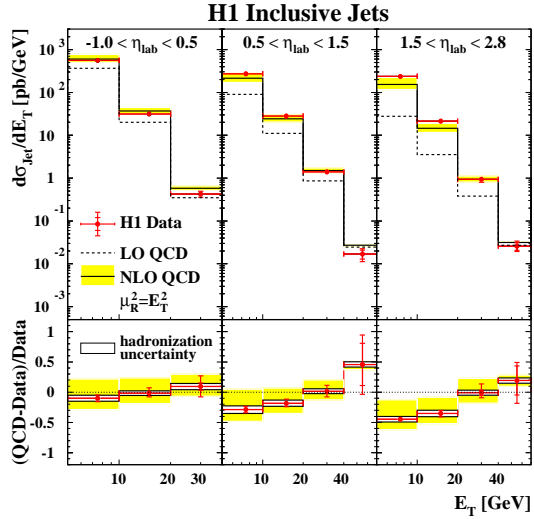}
\caption{\small Inclusive jet cross section in DIS in the Breit system integrated over
$5<Q^2<100$ GeV$^2$ and $0.2 <\eta_{\rm {lab}} < 0.6$, NLO QCD: DISENT without
hadronization correction; shaded band: change of renorm. scale by 1/2 and 2
(fig. from \cite{ref18})}
\label{fig3}
\end{figure}

H1 has extended the inclusive jet measurements \cite{ref8} to low $Q^2$ values
of $5 < Q^2 < 100$ GeV$^2$, $-1 < \eta_{\rm lab} < 2.8$, for $E_T > 5$ GeV and
$0.2 < y < 0.7$. The good agreement of the differential cross section in $E_T$
with NLO QCD (setting $\mu_R = E_T$) in the backward and central region worsens
towards the forward direction i.e. $\eta_{\rm lab} > 1.5$ and for $E_T < 20$
GeV (fig.~\ref{fig3}). In this region, NLO corrections are large and
theoretical uncertainties due to the uncertainty in the renormalization scale
are larger than the experimental errors. Further studies have shown the
discrepancies to NLO QCD to develop at $Q^2 < 20$ GeV$^2$.

\begin{figure} 
\center 
\includegraphics[height=6cm]{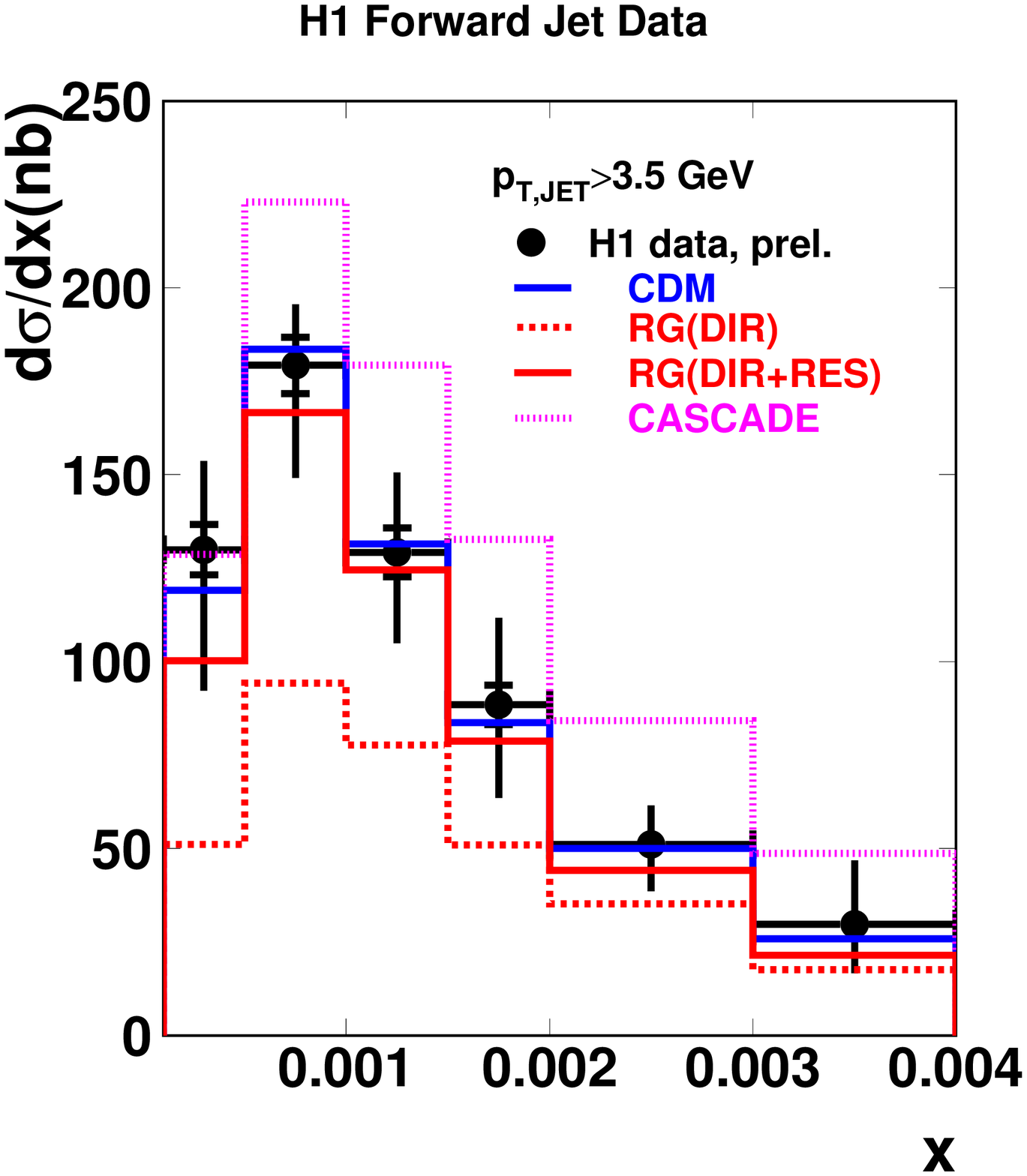}
\includegraphics[height=6cm]{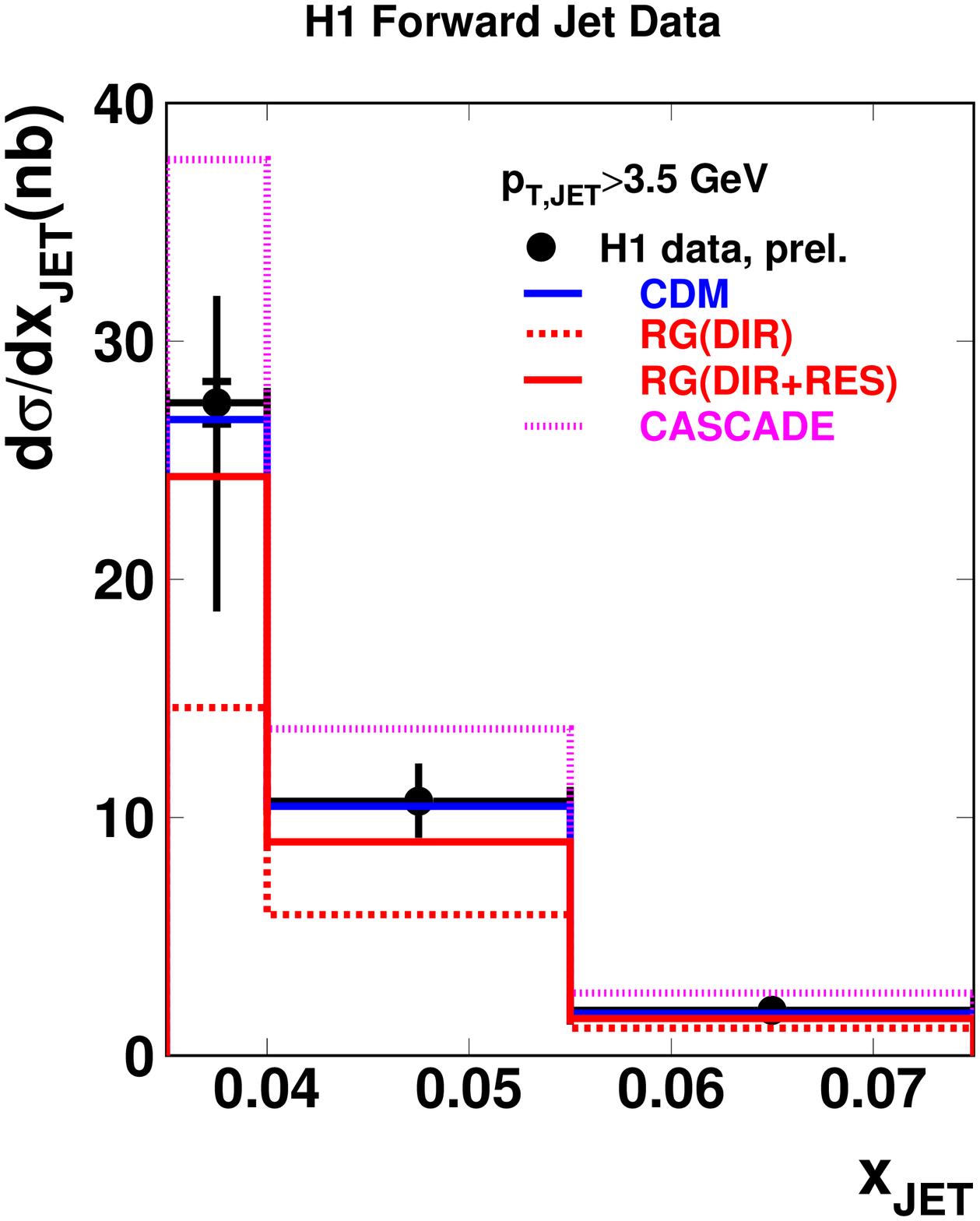} 
\caption{\small Inclusive jet cross section
in DIS in the forward direction i.e. $7^\circ < \theta_{j \rm lab} < 20^\circ$;
CDM: Color-Dipole-Model; RG: RAPGAP; CASCADE: CCFM; (prelim. data, from
\cite{ref9})}
\label{fig4} 
\end{figure}

\subparagraph{(ii)} Data for inclusive jet production near the forward
direction i.e. $7^\circ < \theta_{j \rm lab} < 20^\circ$ have been taken by H1
\cite{ref9} for $0.5 < Q^2/E_T^2 < 2$. The jet analysis using the $k_{\rm
T}$-cluster algorithm has been performed in the lab frame with cuts
corresponding to \mbox{$5<Q^2<75$ GeV$^2$} and $0.1 < y < 0.7$. In
fig.~\ref{fig4} the data are compared with different QCD models. Reasonable
agreement with DGLAP is achieved only if resolved photon processes are
included, but also the colour dipole model describes the data reasonably well.

A process related to forward jet production is forward $\pi^\circ$-production,
which has been studied by H1 \cite{ref10}; here the forward parton is tagged by
a single energetic fragmentation product i.e. the $\pi^\circ$. While in
principle smaller angles than in jet production can be reached, the cross
sections are lower and the hadronization uncertainties are higher. The data can
be described by DGLAP models only after including resolved photon processes,
but also BFKL models provide a reasonable description of the data.

\subparagraph{(iii)} LO contributions to dijet production are the QCD-Compton
effect (QCDC) and boson gluon fusion (BGF) (see fig.~\ref{fig1}). In the high
$Q^2$ region where QCDC is dominating and the pdfs are well constrained by fits
to $F_2$ data, dijet measurements using $\alpha_s$ as input can serve to test
pQCD; alternatively, by comparing the data with pQCD, $\alpha_s$ (and its
running) can be determined. At low $Q^2$, where the cross section is dominated
by BGF, the comparison of data with assumptions on the pdfs can constrain the
gluon density distribution.

\begin{figure}
\center
\includegraphics[height=8cm]{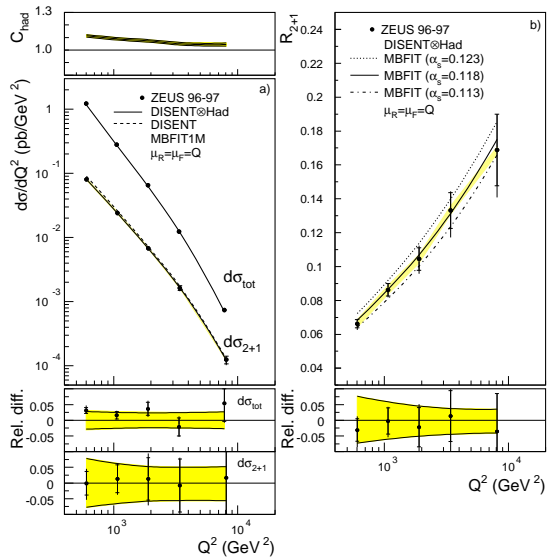}
\caption{\small Inclusive jet $(d\sigma_{tot})$ and dijet $(d\sigma_{2+1})$ cross
sections with dijet fraction $R_{2+1}$ in DIS; shaded band: uncertainty in
absolute energy scale of jets (fig. from \cite{ref11})}
\label{fig5}
\end{figure}

Dijet data have been taken by H1 \cite{ref1} and ZEUS \cite{ref11} in similar
kinematical regions and analyzed by similar methods. Asymmetric cuts have been
applied in $E_T$ in order to avoid infrared sensitive regions of the phase
space. At not too low $Q^2$ and $E_T$ i.e. $Q^2 > 10$ GeV$^2$ and $E_T
> 5$ GeV, the data are reproduced (fig.~\ref{fig5}) within 10\% by NLO QCD
(DISENT). For the $\alpha_s$ analysis of these data see Sect.~\ref{sec:5}.

\begin{figure}
\center
\includegraphics[height=9cm]{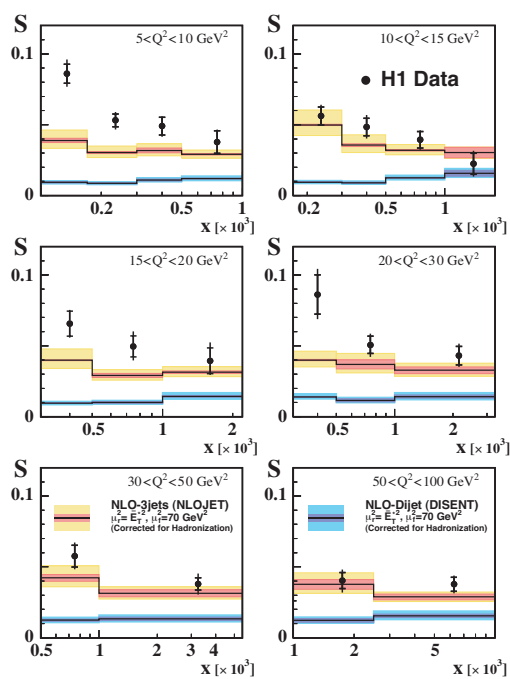}
\caption{\small Ratio $S$ of number of dijet events in DIS with azimuthal
sep. $\phi < 2\pi/3$  between the two highest transverse momentum jets to
the total number of dijet events compared with LO and NLO QCD predictions for
dijet and trijet production; light shaded band: quadr. sum of hadroniz. error
(dark shaded band) and renorm. error; (fig. from \cite{ref12})}
\label{fig6}
\end{figure}

Recent dijet-data from H1 \cite{ref12} in the low-$x$ region i.e. $10^{-4} < x
< 10^{-2}$ for $10 < Q^2 < 100$ GeV$^2$ agree within errors with DGLAP based
NLO QCD predictions for inclusive cross sections. Serious disagreement is
observed, however, in the ratio $S$ of the number of dijet events with
azimuthal separation $\phi < 2\pi/3$ to the total number of dijet events. While
the LO (DISENT) prediction is much to low, the agreement is improved except in
the low-$x$ and low-$Q^2$ region if a third hard jet is included in the NLO
(NLOJET) calculation (fig.~\ref{fig6}). A comparison with RAPGAP shows that the
inclusion of resolved processes improves the agreement except in the very 
low-$x$ and low-$Q^2$ region. From a comparison of the data with the CCFM based
CASCADE MC a strong sensitivity to the assumptions on the unintegrated gluon
distribution can be inferred, which may be used to constrain the gluon pdf.
ARIADNE, which is based on CDM, describes the low-$x$ and -$Q^2$ region
reasonably well, fails, however, at larger $Q^2$ values.

The jet studies in deep inelastic scattering have been extended to trijets
\cite{ref13}, \cite{ref14} as well as to subjets \cite{ref15}.

\section{Jets in photoproduction}
\label{sec:3}

The description of photoproduction involves the convolution of the hard
partonic scattering cross section with the pdfs of both the proton and the
photon. In standard jet production i.e. jets from light quarks only, for the
energy scale $\mu_R = \mu_F = \mu = E_T$ is taken. Apart from the folding with
the electron bremsstrahlung spectrum, the jet cross section is written as

\[
\sigma_j = \sum_{a,b}\int \int dx_\gamma dx_p f_{pa} (x_p,
\mu^2)f_{\gamma b}(x_\gamma,\mu^2)d\sigma_{ab}
(x_p,x_\gamma, \alpha_s,\mu^2)(1+\delta_{\rm{hadr}}),
\]

\noindent where the sum is to be taken over all partons $a, b$ from the proton
and photon.

\subparagraph{(i)} In single jet production a wider kinematic range is
accessible than in dijet production, cross sections are higher and systematic
errors due to the treatment of infrared unsafe regions are avoided; on the
other hand, the reaction is less sensitive to details of the hard scattering
process.

\begin{figure} 
\center 
\includegraphics[height=6cm]{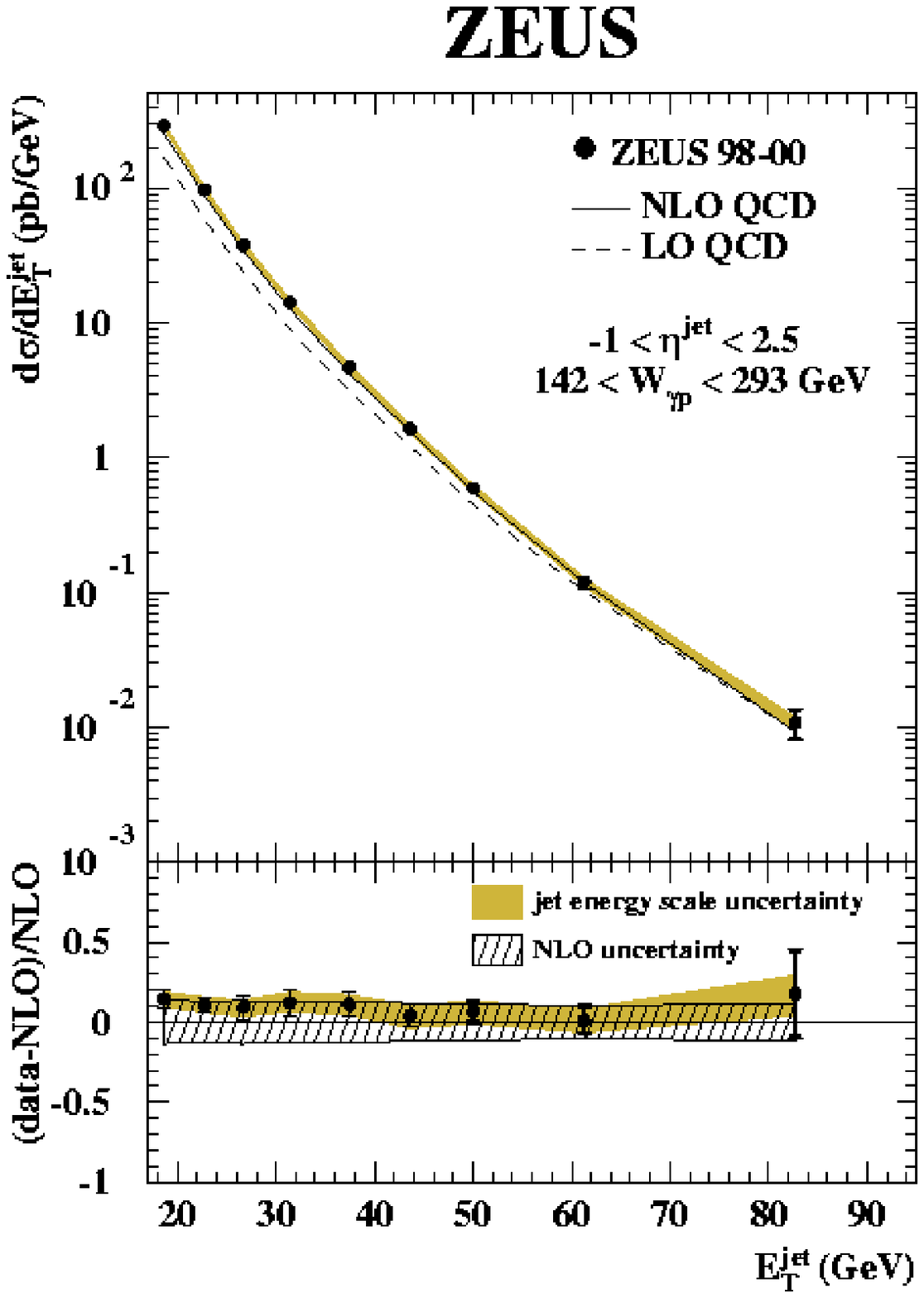} 
\includegraphics[height=6cm]{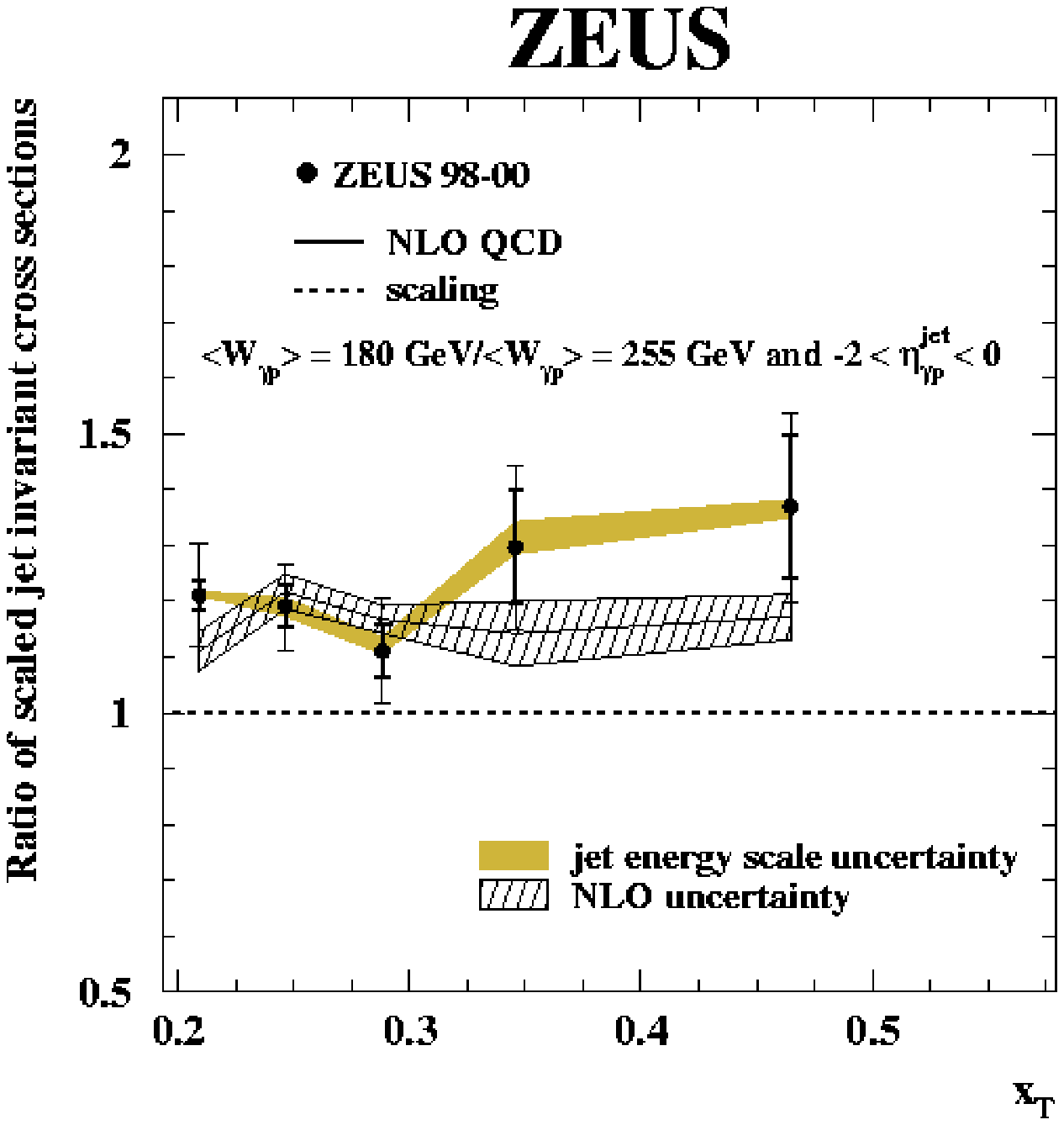}
\caption{\small Inclusive jet production cross section in photoproduction. Ratio of
scaled invariant cross sections (av. over $-2 < \eta_{\rm lab} < 0$) for the two shown
\mbox{$\langle W_{\gamma p} \rangle $} compared with NLO QCD; (fig. from
\cite{ref16})}
\label{fig7}
\end{figure}

In recent measurements at $Q^2 < 1$ GeV$^2$ of differential cross sections by
ZEUS \cite{ref16} and H1 \cite{ref17} using the $k_{\rm T}$-cluster algorithm
in the lab frame, the kinematical range of $E_T > 5$ GeV, $-1 < \eta_{\rm lab}
< 2.5$ and the photon-proton-cms energy $W$ of $95 < W < 293$ GeV has been
covered. While LO QCD calculations fail to reproduce the data in their
normalization, good agreement is found with the corresponding NLO predictions
(fig.~\ref{fig7}, left).  A subset of data taken by H1 at $Q^2 < 0.01$ GeV$^2$
for $5< E_T < 12$ GeV indicate an excess over predictions towards increasing
$\eta$ values. The precision of the experimental data and of the predictions is
not yet sufficient to discriminate between different proton and photon pdfs.
For the determination of $\alpha_s (M_Z)$ and its scale dependence from the
data of \cite{ref12} see Sect.~\ref{sec:5}.

In the ZEUS experiment \cite{ref16} also the scaled invariant jet cross section
\linebreak[4] $E_T^4 Ed^3\sigma/dp_xdp_ydp_z$ has been measured as a function
of the dimensionless variable $x_{\rm T} = 2E_T/W_{\gamma p}$ at $\langle
W_{\gamma p} \rangle = 180$ and 255 GeV. The cross section ratio as function of
$x_T$ violates scaling as predicted from QCD (fig.~\ref{fig7}, right).

\subparagraph{(ii)} The dijet photoproduction cross section is dependent on the
dijet angle $\theta^*$ in the parton-parton-cms, which is sensitive to the
matrix element of the hard scattering subprocess. Direct photon processes are
dominated by BGF with the angular dependence of the spin 1/2-progator $\sim
(1-\cos \theta^*)^{-1}$ whereas in resolved processes gluon-exchange with the 
steeper angular dependence of the spin 1-progator $\sim (1-\cos \theta^*)^{-2}$
is dominating. The dijet events can be enriched with direct or resolved
processes by cutting on the fraction $x^{obs}_\gamma$ of the photon-momentum
transferred to the two jets of highest transverse energy, which is given by
$x^{obs}_\gamma = (E_{T1} e^{-\eta_1}+ E_{T2} e^{-\eta_2})/2yE_e$ with $y$ the
fractional electron energy carried by the photon in the $p$-rest frame.

\begin{figure}  
\center  
\includegraphics[height=8cm]{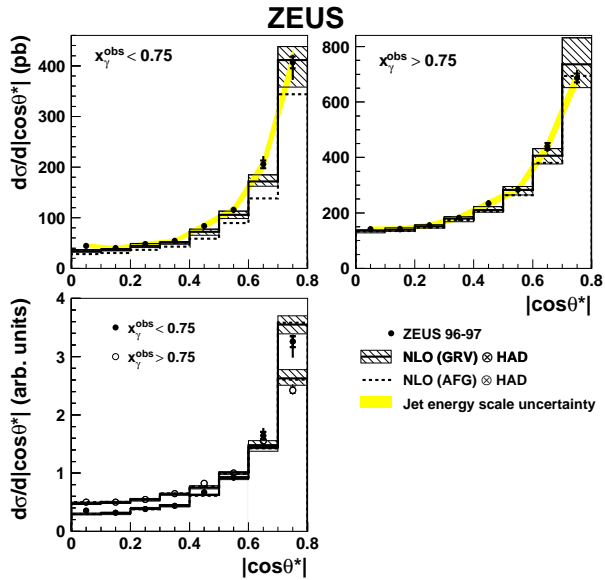}
\caption{\small Inclusive dijet photoproduction as a function of the dijet angle
$\theta^*$ in the parton-parton-cms for different cuts on the fractional photon
momentum $x_\gamma^{obs}$ transferred to the dijet system. Comparison with NLO QCD
for different photon-pdfs; (fig. from \cite{ref18})}
\label{fig8}  
\end{figure}

Dijet photoproduction has been studied by ZEUS \cite{ref18} and H1 \cite{ref19}
under similar kinematical conditions using the $k_{\rm T}$-cluster algorithm.
The differential cross sections in the dijet invariant mass, $E_T$ and $\eta$
agree well with NLO QCD expectations and in particular the measured dijet
angular distributions (fig.~\ref{fig8}) shows the expected difference for
direct $(x^{obs}_\gamma > .75)$ and resolved processes $(x^{obs}_\gamma < .75)$. The
sensitivity of the data to the parametrization of the photon-pdfs appears to be
affected by the choice for the $E_T$-cuts (ZEUS: $E_{T1} > 14$ GeV, $E_{T2} >
11$ GeV; H1: $E_{T1} > 25$ GeV, $E_{T2} > 15$ GeV).

\section{$\alpha_s$-determinations}
\label{sec:4}

The good agreement of jet cross sections with QCD predictions has enabled
determinations of $\alpha_s$ and its scale dependence.

The H1 inclusive jet cross sections \cite{ref5} as a function of $E_T$ have
been fitted to the QCD prediction for four regions of $Q^2$ using the
proton-pdfs and $\mu_R$ and $\mu_F$ as input. The resulting $\alpha_s$ has been
shown to be stable against variations of the jet algorithm. The combined fit to
all $Q^2$-data evolved to $m_Z$ is shown in the summary {fig.~\ref{fig9}).

\begin{figure}
\center
\includegraphics[height=8cm]{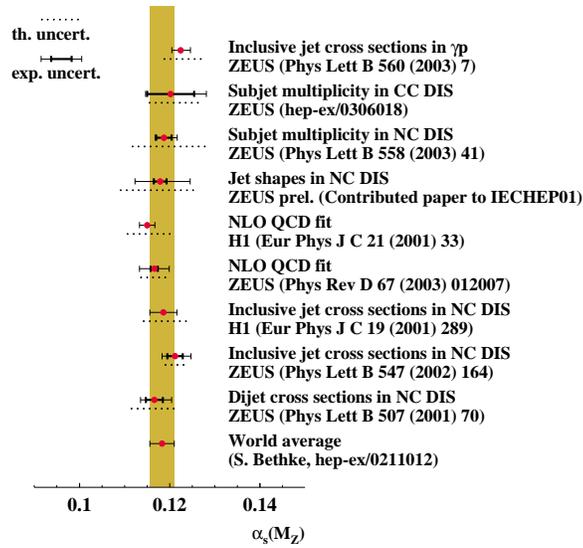}
\caption{\small $\alpha_s(M_Z)$ from recent H1 and ZEUS measurements and the world
average}
\label{fig9}
\end{figure}

For the analysis of the ZEUS data \cite{ref6} the following procedure has been
applied: The cross sections $d\sigma/dA$ with $A=Q^2$, $E_T$ have been
calculated in NLO QCD for the same pdf-set with three $\alpha_s$-values. These
calculations were used to parametrize the $\alpha_s$-dependence of the binned
cross sections $(d\sigma/dA)_i$ for each bin $i$ according to
$(d\sigma(\alpha_s)/dA)_i = C_{1i}.$ $\alpha_s^1 + C_{2i}$. $\alpha_s^2$ with
$\alpha_s = \alpha_s(M_Z)$. From a $\chi^2$-fit of this ansatz to the measured
$d\sigma/dA$, $\alpha_s$ was obtained for the chosen regions of $A$. The best
fit was obtained for $Q^2 > 500$ GeV$^2$ (fig.~\ref{fig9}). The same procedure
has been applied to determine the scale dependence of $\alpha_s$. With $E_T$ as
energy scale, the $\alpha_s$ dependence of $d\sigma/dE_T$ was parametrized in
terms of $\alpha_s(\langle E_T \rangle)$ where $\langle E_T \rangle$ is the
mean value of $E_T$ in bin $i$. The result is shown in fig.~\ref{fig10}.

\begin{figure}
\center
\includegraphics[height=7cm]{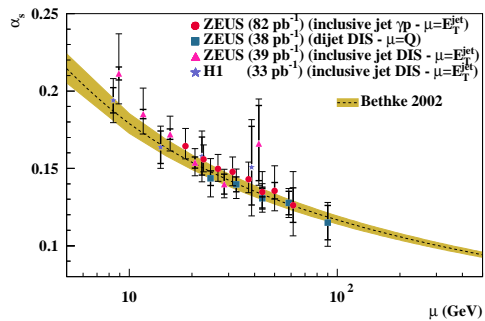}
\caption{\small Scale dependence of $\alpha_s$ from recent H1 and ZEUS measurements and QCD
predictions for $\alpha_s(M_Z) = 0.1183$ (world average)}
\label{fig10}
\end{figure}

The same method has been applied to other jet results from ZEUS i.e. the
dijet fraction \cite{ref12}, subjet multiplicities (\cite{ref15}) and jets from
photoproduction \cite{ref15}. A summary of the results is shown in figs.
\ref{fig9} and \ref{fig10}.

\section{Summary}
\label{sec:5}

In the analysis of jet production in DIS and photoproduction at HERA the
longitudinally invariant $k_{\rm T}$-cluster algorithm in its inclusive version
has become the standard jet finding algorithm.

In DIS, the inclusive jet cross sections at higher values of $Q^2$ and $E_T$ are
well described by NLO QCD; in this region higher order and hadronization
corrections are small; this holds as well for photoproduction.

\clearpage 

In the forward region and/or at smaller $Q^2$ resp. $x$ values, the situation
is less satisfactory. DGLAP based calculations are expected to become less
reliable in this region, the mentioned corrections are sizable and the hadronic
structure of the photon i.e. resolved processes have to be taken into account.
Calculations based on BFKL or CCFM evolution only partly show better agreement.
An increase in experimental statistics and efforts to reduce the theoretical
uncertainties are highly desirable for a better understanding of this
challenging region.

The measurements of $\alpha_s$ and its running from jet data have yielded
results which are of perhaps unexpected precision; they can well compete with
other precision measurements e.g. from $e^+e^-$-annihilation and are in good
agreement with the world average.

\vspace{0,5cm}

\underline {Acknowledgements}
I thank G. Grindhammer for valuable discussions and comments on the paper.


\end{document}